\documentstyle[12pt,aaspp4]{article}

\newcommand{\eg}{{\it e.g.}}
\newcommand{\ie}{{\it i.e.}}
\newcommand{\etal}{~{\it et~al.}\ }  
\newcommand{\ho}{$H_0$}
\newcommand{\prhchi}{PRH$\chi^2$}

\newcommand{\unit}[1]{\ifmmode\,{\rm #1}\else$\,{\rm #1}$\fi}
\newcommand{\kmsmpc}{\unit{km\,s^{-1}Mpc^{-1}}}
\newcommand{\cm}{\unit{cm}}

\newcommand{\Bjet}{B_{\rm jet}}
\newcommand{\Bcore}{B_{\rm core}}
\newcommand{\Ajet}{A_{\rm jet}}
\newcommand{\Acore}{A_{\rm core}}
\newcommand{\Rjet}{R_{\rm jet}}
\newcommand{\Rcore}{R_{\rm core}}
\newcommand{\Rvla}{R_{\rm VLA}}

\newcommand{\RAC}{R_{\rm AC}}
\newcommand{\BDC}{B_{\rm DC}}
\newcommand{\BAC}{B_{\rm AC}}
\newcommand{\ADC}{A_{\rm DC}}

\newcommand{\BcoreDC}{B_{\rm core,DC}}
\newcommand{\AcoreDC}{A_{\rm core,DC}}

\lefthead{Haarsma\etal}
\righthead{Radio Time Delay of Lens 0957+561}
\begin{document}

\title{The Radio Wavelength Time Delay of \\ 
	Gravitational Lens 0957+561 }
 
\author{D. B. Haarsma\altaffilmark{1,}\altaffilmark{2}, 
        J. N. Hewitt\altaffilmark{2}, 
        J. Leh\'ar\altaffilmark{3}, 
        and B. F. Burke\altaffilmark{2}}
 
\altaffiltext{1}{Haverford College, Haverford, PA 19041, 
dhaarsma@haverford.edu}
 
\altaffiltext{2}{Department of Physics, 37-607, Massachusetts 
Institute of Technology, Cambridge, MA 02139}
 
\altaffiltext{3}{Harvard-Smithsonian Center for Astrophysics, 
60 Garden Street, Cambridge, MA 02138}

%%%%%%%%%%%%%%%%%%%%%%%%%%%%%%%%%%%%%%%%%%%%%%%%%%%%%%%%%%%%%%%%%%%%%%%%
\begin{abstract}

The gravitational lens 0957+561 was monitored with the Very Large
Array from 1979 to 1997. The 6~cm light curve data from 1995-1997 and
the 4~cm data from 1990-1997 are reported here.  At 4~cm, the
intrinsic source variations occur earlier and are twice as large as
the corresponding variations at 6~cm.  The VLBI core and jet
components have different magnification factors, leading to different
flux ratios for the varying and non-varying portions of the VLA light
curves.  Using both the PRH$Q$ and Dispersion statistical techniques,
we determined the time delay, core flux ratio, and excess non-varying
B~image flux density.  The fits were performed for the 4~cm and 6~cm
light curves, both individually and jointly, and we used Gaussian
Monte Carlo data to estimate 68\% statistical confidence levels.  The
delay estimates from each individual wavelength were inconsistent
given the formal uncertainties, suggesting that there are unmodeled
systematic errors in the analysis.  We roughly estimate the systematic
uncertainty in the joint result from the difference between the 6~cm
and 4~cm results, giving $409\pm30$~days for the PRH$Q$ statistic and
$397\pm20$~days for the Dispersion statistic.  These results are
consistent with the current optical time delay of $417\pm3$~days,
reconciling the long-standing difference between the optical and radio
light curves and between different statistical analyses.  The
unmodeled systematic effects may also corrupt light curves for other
lenses, and we caution that multiple events at multiple wavelengths
may be necessary to determine an accurate delay in any lens system.
Now that consensus has been reached regarding the time delay in the
0957+561 system, the most pressing issue remaining for determining \ho\
is a full understanding of the mass distribution in the lens.

\end{abstract}
 

%%%%%%%%%%%%%%%%%%%%%%%%%%%%%%%%%%%%%%%%%%%%%%%%%%%%%%%%%%%%%%%%%%%%%%%%%%%
 
\section{Introduction}
\label{intro}

The time delay between multiple gravitationally lensed images can be
used to measure the distance of high redshift objects, and thus is a
useful estimator of the Hubble parameter, \ho.  After many years of
monitoring the lens 0957+561, the time delay estimates for this system
are finally converging on an accepted value.  Groups monitoring
0957+561 at optical wavelengths have detected a sharp variation in
each image, and have found the optical delay to be $417\pm3~$days
(\cite{kundic95a}, 1997;
%\nocite{kundic97a}
\cite{oscoz97a}; \cite{schild97a}).  Given the long controversy over
the value of the delay (for a history, see Table 1 of Haarsma\etal 1997,
%\nocite{haarsma97a}
hereafter Paper~1), it is important that the optical measurement be
confirmed at radio wavelengths.  The MIT radio astronomy group has
monitored the source at radio wavelengths from 1979 to 1997 and the
final light curve data and time delay results are reported here.

%%%%%%%%%%%%%%%%%%%%%%%%%%%%%%%%%%%%%%%%%%%%%%%%%%%%%%%%%%%%%%%%%%%%%%%%
\section{Observations}
\label{obs}

Observations have occurred monthly at the National Radio Astronomy
Observatory (NRAO) Very Large Array radio telescope (VLA)\footnote{The
National Radio Astronomy Observatory is operated by Associated
Universities, Inc., under cooperative agreement with the National
Science Foundation.} since 1979 at 6~cm and since 1990 at 4~cm.  The
monitoring ended in December~1997.  All of the data were reduced in
the manner described in Paper~1 and Leh\'ar\etal(1992).
%\nocite{lehar92a}
To determine the flux densities of the point images, it was necessary
to subtract the extended structure in the field.  At both 6 and 4~cm,
this subtraction was difficult in the most compact VLA array, D, thus
there are gaps in the light curves for 4 months of every 16 month
cycle.  In addition, some observations in other VLA arrays were
excluded due to bad weather or poor subtraction of the extended
structure.  When the observations were made in a combination or
non-standard array configuration, the data were analyzed according to
the next largest standard configuration (A, B, or C).

The 6~cm data through December~1994 were presented in Paper~1.  The
remaining 6~cm data and all of the 4~cm data are given in Tables~1 and
2, and plotted in Figure~1.  The light curve data are also available
electronically through {\tt http://space.mit.edu/RADIO/papers.html}.
There are a total of 147 points in the 6~cm light curve, and 58 points
in the 4~cm light curve.  At 6~cm, the flux density of the B image
increased in 1995, following the A image increase in 1994.  The
current 6~cm feature has lasted longer than the similar feature around
1989-1991, but the A image is now declining.  At 4~cm, the quasar is
twice as variable as at 6~cm (as a percentage of average flux
density).  Also, the variations in the 4~cm light curves occur earlier
than the corresponding features at 6~cm.  Both of these
characteristics are consistent with multi-wavelength models and other
observations of AGN variability (\eg\ \cite{marscher85a};
\cite{stevens96a}).  The well-sampled increase and decrease at 4~cm in
1994-97 has helped significantly in determining the radio time delay.

%%%%%%%%%%%%%%%%%%%%%%%%%%%%%%%%%%%%%%%%%%%%%%%%%%%%%%%%%%%%%%%%%%%%%%%%%
\section{Free parameters in the Light Curves}
\label{parameters}

When fitting for the time delay between the images, the difference in
magnification between them must be properly taken into account.  In
past analyses of lensed light curves (including Paper~1), only two
parameters were used in the fit: the time delay and a single flux
ratio.  Conner, Leh\'ar, \&~Burke~(1992),
%\nocite{conner92a}
however, have pointed out that the magnification varies rapidly along
the B image, causing the VLBI core and jet components to have
different flux ratios, with the core ratio being larger.  At the
resolution of the VLA, the beam includes both the core and the jet,
and thus the flux ratio of the VLA light curves $\Rvla$ is a composite
of the core and jet values.  The VLA light curve is the sum of the jet
(which is constant in time), and the core (which has both variable and
constant components).  There are then four physical parameters: the
time delay $\tau$, the flux ratio of the core $\Rcore =
\Bcore/\Acore$, the flux ratio of the jet $\Rjet = \Bjet/\Ajet$, and
the amount of flux density due to the jet vs. the core, \ie\ $B(t) =
\Bcore(t) + \Bjet$ for the B image.  Note that the core can contain
both constant (DC) and variable (AC) components, \ie\ $\Bcore =
\BcoreDC + \BAC$. Also, the DC part of the light curve is due to both
the core and the jet, \ie\ $\BDC = \BcoreDC + \Bjet$.

Press~\&~Rybicki (1998)
%\nocite{press98a}
discuss these issues in the context of the optical light curves of
0957+561.  They point out that the amount of constant flux due to the
core ($\BcoreDC$) is impossible to determine, since we may have not
yet seen the variable part of the core ($\BAC$) go to zero.  They show
that on a fundamental level there are only three measureable
parameters in a pair of lensed light curves, which may be cast as: the
time delay $\tau$, the core flux ratio $\Rcore$, and the extra
constant flux in the B image that does not occur in the A image,
\begin{equation}
	c = \frac{ \BDC }{ \RAC } - \ADC
\end{equation}
(\cite{press98a}), where $\RAC$ is the flux ratio of the variable
component.  It is useful to write $c$ in terms of the core and jet
components of the radio images as
\begin{equation}
	c = \frac{ \Bjet + \BcoreDC }{ \Rcore }
		 - \left( \Ajet + \AcoreDC \right),
\end{equation}
and therefore
\begin{equation}
	c = \Bjet \left( \frac{1}{\Rcore} - \frac{1}{\Rjet} \right),
\label{eq.c0}
\end{equation}
where the DC core components cancel out.  The value of $c$ can thus be
estimated from the values of $\Bjet$, $\Rcore$, and $\Rjet$.  Since in
the case of 0957+561 we have $\Rjet < \Rcore$ (\cite{conner92a}), the
value of $c$ must be negative; \ie\, the A curve has a larger amount
of constant flux than the core-ratio corrected B curve.

The values of several of the above parameters can be estimated from
observations without doing time delay fitting. Garrett\etal~(1994)
%\nocite{garrett94a}
compiled the information on the core flux ratio from VLBI and optical
observations, and found the weighted average of these estimates to be
$\Rcore = 0.75\pm0.02$.  Also, the faintest portions of the VLA light
curves set upper limits on the jet flux density, i.e. $\Bjet \lesssim
21$~mJy at 6~cm, and $\Bjet \lesssim 15$~mJy at 4~cm.  A better
estimate of $\Bjet$ can be obtained by comparing coincident VLBI and
VLA observations.  The VLBI observations give the core flux density at
a particular epoch, which can be subtracted from the VLA flux density
to obtain the VLA jet flux density.  Campbell\etal~(1995)
%\nocite{campbell95a}
report VLBI observations at 6~cm on 1987~Sep~28 and 1989~Sep~26, and
by comparing these to VLA observations occurring on the same days we
find $\Bjet=11.1\pm0.4$~mJy and $\Rjet = 0.63\pm0.03$.  The values for
$\Rcore$, $\Rjet$, and $\Bcore$ can be combined using
equation~\ref{eq.c0} to find $c_6= -2.7\pm0.8$~mJy.

The above estimates are all for the 6~cm light curves.  At 4~cm there
are no coincident VLBI/VLA observations, so we can not make similar
estimates.  The value of $c$ is different at 6~cm and 4~cm due to the
difference in $\Bjet$; note that the ratios $\Rjet$ and $\Rcore$ are
the same for the two bands.  For a synchrotron spectrum, $\Bjet$ will
be smaller at 6~cm than 4~cm, and thus we expect $|c_4|$ to be smaller
than $|c_6|$.

%%%%%%%%%%%%%%%%%%%%%%%%%%%%%%%%%%%%%%%%%%%%%%%%%%%%%%%%%%%%%%%%%%%%%%%%%
\section{Time Delay Analysis Methods}
\label{methods}

To fit for the three parameters $\tau$, $\Rcore$, and $c$ (described
in \S\ref{parameters}), we used the PRH$Q$ statistic (Press, Rybicki,
\&~Hewitt 1992a, 1992b;
%\nocite{press92a} \nocite{press92b}
\cite{rybicki92a}; incorporating the modifications of
\cite{rybicki94a}; \cite{press98a}), and the Dispersion statistic
(\cite{pelt94a}, 1996),
%\nocite{pelt96a}
which were described in Paper~1.  We used linear units (mJy) rather
than the logarithmic units defined in Paper~1. The discrete
correlation function (\cite{lehar92a}) did not find a strong
correlation in the 4~cm light curves, so that statistic was not used
here. Gaussian Monte Carlo data were made as described in Paper~1, but
now with the four physical parameters $\tau$, $\Rcore$, $\Rjet$, and
$\Bjet$.  Five hundred Gaussian Monte Carlo data sets were used to
estimate the 68\% confidence intervals on the results for the real
light curves, where the fitted $c$ values were compared to the input
parameters using equation~\ref{eq.c0}.  The pseudo-jackknife test from
Paper~1 was used to test the stability of the result to the removal of
individual points.

To determine whether neglecting the difference between the core and
jet flux ratio caused an error in our previous analysis, we applied
the two dimensional fit (for $\tau$ and $\Rvla$, as in Paper~1) to the
Gaussian Monte Carlo data made with four parameters.  The resulting
fitted-minus-true values did not show a significant bias (to long or
short delays, for example), but did show an increase in scatter about
the true delay.  We found that the error in the delay increased
monotonically with $\Bjet$, from roughly 20~days for $\Bjet=0$~mJy to
roughly 100~days for $\Bjet\sim11$~mJy (the value for the real light
curves) when using the PRH$Q$ statistic and the 6~cm Monte Carlo data.
The same test with the 4~cm Monte Carlo data, and with the Dispersion
statistic at both 4~cm and 6~cm, revealed a similar but somewhat
milder effect, with the delay error at least doubling between small
and large values of $\Bjet$.  This dependence on the amount of flux
density in the jet component may be one cause of the inconsistency in
delay estimates over the years, and we caution that fitting for only
two parameters may introduce significant errors.

In addition to analyzing the two wavelengths individually, we also
fitted for the parameters using both wavelengths at once.  The
Dispersion and PRH$Q$ statistics are both easily modified for this by
minimizing the sum of the statistics from each wavelength (see
Press\etal 1992b), 
%\nocite{press92b} 
and fitting for the parameters $\tau$, $\Rcore$, $c_6$, and $c_4$.
Monte Carlo analysis was also done for the joint data, where the 6~cm
and 4~cm Monte Carlo sets were constructed with the same set of
[$\tau$, $\Rcore$, $\Rjet$, $\Bjet$].

The covariance model (Paper 1; Press\etal 1992a)
%\nocite{press92a}
used for the PRH$Q$
statistic was found by an iterative procedure on the individual light
curves.  First, measurement errors of 2\% were assumed and used to
make point estimates for the structure function, and then fitted to an
exponential in the lag range of 100 to 700 days.  This structure
function was then used to determine the \prhchi\ value for the light
curve, and the measurement errors were adjusted until \prhchi\ equaled
the degrees of freedom.  Then the process was repeated for the new
measurement error value.  Iterations stopped when the square root of
(\prhchi/degrees of freedom) changed by less than 1\% when \prhchi\
was calculated with the measurement error of the previous iteration.
At 6~cm, the covariance model found was
\begin{equation}
	V(T) = 1.673\times10^{-4} T^{1.606}\unit{mJy}^2,
\label{eq.cstruct}
\end{equation}
with measurement errors $e_A=1.82\%$ and $e_B=2.34\%$, where $T$ is
the time lag between two points on the curve. At 4~cm, the fitted
covariance model was
\begin{equation}
	V(T) = 3.174\times10^{-4} T^{1.633}\unit{mJy}^2,
\label{eq.xstruct}
\end{equation}
with measurement errors $e_A=1.67\%$ and $e_B =2.19\%$.

The sharp feature in the B image at 6~cm in Spring~1990 is
statistically inconsistent with the rest of the light curve (see Paper
1), and, given the short delay found from the optical and 4~cm light
curves, the A image shows that the feature is not intrinsic to the
source.  Thus we expect that the results with the points removed will
be more accurate than the results for the full 6~cm curves.  The 6~cm
analysis was done both with and without the four points
(1990~March~15, April~10, May~7, and May~23); the light curve without
the points will be denoted $6^*$cm.

%%%%%%%%%%%%%%%%%%%%%%%%%%%%%%%%%%%%%%%%%%%%%%%%%%%%%%%%%%%%%%%%%%%%%%%%% 
\section{Results}
\label{results}

The main results of the time delay analysis are shown in Tables~3 and
4.  These and other aspects of the results are worth discussion in
this section.  First, the Monte Carlo analysis showed that the use of
the three parameter fit ($\tau$, $\Rcore$, $c$) made the time delay
confidence interval independent of $\Bjet$ (rather than increase with
$\Bjet$ as happened in the two parameter fit described in
\S\ref{methods}).  Next, the pseudo-jackknife test (see Paper~1),
using either the PRH$Q$ or Dispersion statistic, showed that removal
of an individual point from the light curve general caused a change in
the delay that was much smaller than the confidence interval, with a
few important points in the curve (typically during rises or falls)
causing a change at about the amount of the confidence interval.
Thus, the delay estimate in all cases is stable under the removal of
individual points.  Note also that the removal of the four Spring~1990
points ($6^*$cm vs. 6~cm in Tables~3 and 4) never caused a change in
the fitted parameters of more than the confidence interval.

The relationship between the values of $c$ and $\Rcore$ for a given
delay is worth pointing out. Figure~2 shows the
\prhchi\ statistic as a function of $c$ and $\Rcore$ for the $6^*$cm
light curves, with the delay fixed at the best fit value.  The surface
is a diagonal trough, with the location of the minimum poorly
constrained along the bottom of the trough.  The trend is such that a
larger value of $\Rcore$ requires a more negative value of $c$, as
expected from equation~\ref{eq.c0}.  The 4~cm light curve has a
similar \prhchi\ surface, and the Dispersion surface shows a similar
but even more pronounced effect.  Thus, if either $\Rcore$ or $c$ is
poorly constrained by the light curves, the other parameter will also
be poorly determined.

There is a significant bias in $c$ and $\Rcore$ for the Dispersion
statistic, which is related to the interdependence of these
parameters.  The fitted-minus-true values from the Dispersion analysis
of the Gaussian Monte Carlo data showed a pronounced asymmetry and
bias in both $c$ and $\Rcore$ (but was nearly symmetric and un-biased
in delay).  Tests showed that the bias in $c$ and $\Rcore$ was
somewhat reduced for $\delta\sim25$~days (where $\delta$ is a
weighting parameter in the Dispersion statistic, see Paper~1).  We
continued to use $\delta=60$~days (as done by \cite{pelt96a}), since
that value had the narrowest distribution for the fitted-minus-true
delays.  Thus, only the time delay results are given in
Table~4.

The PRH$Q$ statistic produced a symmetric distribution in $c$ and
$\Rcore$, and the results are listed in Table~3.
Note that the fitted value of $c$ is more negative when $\Rcore$ is
larger, as mentioned above.  The analysis of the individual wavelengths
found somewhat different values of $\Rcore$, and that $c_4$ is more
negative than $c_6$ contrary to the predictions of \S\ref{parameters}.
The joint analysis of the two wavelengths, however, used all of the
available information to constrain the fit, and the resulting
$\Rcore$, $c_6$, and $c_4$ are in good agreement with the predictions
of \S\ref{parameters}.

Turning now to the results for the delay, it is of interest that all
of the delay estimates in Tables~3 and 4 are much smaller than the
value of 540~days found by Press\etal(1992b)
%\nocite{press92b}
using the first 80 points in the 6~cm light curve. As explained in
Paper~1, this change in the delay estimate is due entirely to the
addition of new features to the light curve.  If the first 80 points
in the curve (with or without the four Spring~1990 points) are fitted
for the three parameters, the Dispersion statistic finds a delay of
roughly $550\pm35$~days, and the PRH$Q$ statistic (using the above
covariance model, eq.~\ref{eq.cstruct}) finds a delay of roughly
$525\pm25$~days (the confidence intervals were estimated from Monte
Carlo analysis of 100 light curves).  Therefore the change in the
delay estimate between the first 80 points and the current 147 points
occurs for both statistical methods, for fits with two or three
parameters, and for a variety of covariance models.  It does not,
however, occur in our Gaussian Monte Carlo data.  Comparing delay
estimates from the first 80 points and the full curves in the Monte
Carlo data, we find that 99\% of the sets have a difference in delay
less than the 75 day difference seen in the real light curves.  Thus
the Monte Carlo light curves must still be missing some characteristic
of the real data.

A related issue is the difference in the delay estimates for the two
wavelengths (for PRH$Q$, $\tau_6=452^{+14}_{-15}$~days vs.
$\tau_4=397\pm12$~days, a difference of about three confidence
intervals).  The time delay between lensed images should be completely
independent of wavelength, and indeed the optical and radio estimates
come remarkably close.  This effect also does not occur in the
Gaussian Monte Carlo data, which were constructed such that the 6 and
4\cm\ data have the same set of [$\tau$, $\Rcore$, $\Rjet$, $\Bjet$].
When applying the PRH$Q$ statistic, we found that 99\% of the Monte
Carlo curves had a difference of ($|\tau_6-\tau_4|$) that was smaller
than the 55~day difference in the real data; similarly, the Dispersion
statistic found that about 80\% of the data sets had a smaller delay
difference.

Since both of these effects (the significant change in the delay
estimate as features are added to the curves, and the significant
difference between the two radio wavelengths) are not seen in the
Monte Carlo data, there must still be some systematic effect that
has not been taken into account in the creation of the Monte Carlo data
or in our analysis.  One source of the systematic error may be
interstellar scintillation, which can cause variability at a level not
much larger than our observational error of 2\%, perhaps creating
small features such as the discrepancy between the 6~cm A and B images
in early 1985 (see Figure~4).  Although individual features of a few
percent are difficult to identify, they may cause a significant bias
in the delay estimate if they occur at crucial times in the light
curves.  We note that microlensing could cause similar low level
systematic effects in the optical light curves (\cite{schild91a};
\cite{schmidt98a}).  Since it is beyond the scope of this paper to
model these systematic effects, we make a rough estimate from the
difference between the 6 and 4\cm\ delay estimates, giving a
systematic uncertainty of roughly $\pm30$~days for PRH$Q$ and
$\pm20$~days for the Dispersion.  Note that this uncertainty is
still less than 10\%, and is only one factor contributing to the error
in \ho\ (see \S6). Note also that the optical time delay estimate has a
smaller error (only 1\%,\cite{kundic97a}) primarily because the
source varies much more rapidly at optical wavelengths than radio
wavelengths.  We caution others monitoring gravitational lenses that
in order to determine an accurate time delay it is preferable to have
light curves with multiple features at multiple wavelengths, since the
estimate of the time delay based on a single feature at a single
wavelength could easily be corrupted by these low level systematic
effects (as the 0957+561 6~cm curves were after the first 80
observations).  The 0957 radio light curves will continue to be a
useful data set for studying systematic effects and time delay
analysis techniques.

Figure~3 shows the PRH$Q$ and Dispersion statistics as a function of
delay for the joint analysis of the 4~cm and $6^*$cm curves, with the
values of $c_6$, $c_4$, and $\Rcore$ fixed at the best fit values.
Figure~4 shows the aligned light curves at the two wavelengths with
the PRH optimal reconstruction (see Press\etal 1992a
%\nocite{press92a} 
and Paper~1).

%%%%%%%%%%%%%%%%%%%%%%%%%%%%%%%%%%%%%%%%%%%%%%%%%%%%%%%%%%%%%%%%%%%%%%%%% 
\section{Conclusions}
\label{conclusions}

Since our last report (Paper~1), the B image has increased at 6~cm,
and the A image has entered a slow decline.  The 4~cm curves, given
here for the first time, are highly variable and give additional
features to aid in determining the time delay.  To take into account
the difference in magnification of the core and jet components of each
image, we have fit for the delay, core flux ratio, and the excess flux
density in the B image (as defined by \cite{press98a}).  The delay
estimates found from the wavelengths individually disagree by a few
confidence intervals, indicating that there are systematic effects not
modeled in our analysis.  The delay estimates found from the joint
analyses of both wavelengths were $409\pm30$~days for PRH$Q$ and
$395\pm20$~days for the Dispersion, where the uncertainty is based on
a rough estimate of the systematic error.  Both results are consistent
with the delay estimated from optical monitoring ($417\pm3$~days,
\cite{kundic97a}), and thus we now have good agreement for the value
of the delay from both statistics in both the radio and optical light
curves.  Consensus has finally been reached on the value of the delay
for gravitational lens 0957+561.

This measurement of the delay can now be used to answer cosmological
questions.  The Hubble parameter, however, depends not only on the
delay but also on the lens model and the galaxy velocity dispersion.
Using the SPLS model of Grogin \&~Narayan (1996a, 1996b),
%\nocite{grogin96a} \nocite{grogin96b}
the recent Keck velocity dispersion measurement of
279\unit{km\,s^{-1}} (\cite{falco97a}), and a time delay of 409~days,
we obtain \ho=67\kmsmpc.  In fitting this model, Grogin \&~Narayan
used the ground-based optical position of the lensing galaxy as a
constraint, rather than the more precise VLBI position. Since then,
the HST optical position of the lensing galaxy has been found to agree
with the VLBI position (\cite{bernstein97a}).  The modelers point out
(\cite{grogin96b}) that if the VLBI position is used as the model
constraint, their model fit is very similar to that of Falco,
Gorenstein, \&~Shapiro (1991);
%\nocite{falco91a}
for the same
delay and velocity dispersion, the Falco\etal model gives
\ho=41\kmsmpc.  Thus, the change in the position of the lensing galaxy
causes a change in the estimate of \ho\ well beyond the statistical
error, apparently in contradiction to the conclusions of Kundi\'c\etal(1997)
%\nocite{kundic97a}
regarding the robustness of the \ho\ determination.  New modeling work
must be done which incorporates the improved galaxy position, the
recent observations of the cluster mass distribution
(\cite{fischer97b}), a careful treatment of the systematic errors in
the velocity dispersion (\cite{romanowsky98a}), and the recently
reported structure at x-ray (\cite{chartas98a}), optical
(\cite{bernstein97a}), and radio (\cite{avruch97a};
\cite{harvanek97a}) wavelengths.  These new observations will allow an
improved fit of the model to the data and provide a more accurate
measure of the Hubble parameter.

%%%%%%%%%%%%%%%%%%%%%%%%%%%%%%%%%%%%%%%%%%%%%%%%%%%%%%%%%%%%%%%%%%%%%%%%%
 
\acknowledgements

We thank the VLA staff for their assistance over the many years of
this monitoring project.  DBH and BFB have been supported in part by
the National Science Foundation.  JNH acknowledges the support of a
David and Lucile Packard Fellowship, a NSF Presidential Young
Investigator Award, and NSF grant AST 96-17028.  JL acknowledges the
support of NSF grant AST 93-03527.

%%%%%%%%%%%%%%%%%%%%%%%%%%%%%%%%%%%%%%%%%%%%%%%%%%%%%%%%%%%%%%%%%%%%%%%%%
\clearpage

%\bibliography{apjmnemonic,radio}
%\bibliographystyle{apj}

%%%%%%%%%%%%%%%%%%%%%%%%%%%%%%%%%%%%%%%%%%%%%%%%%%%%%%%%%%%%%%%%%%%%%%%%%%
 
\clearpage

\begin{figure}
\plotone{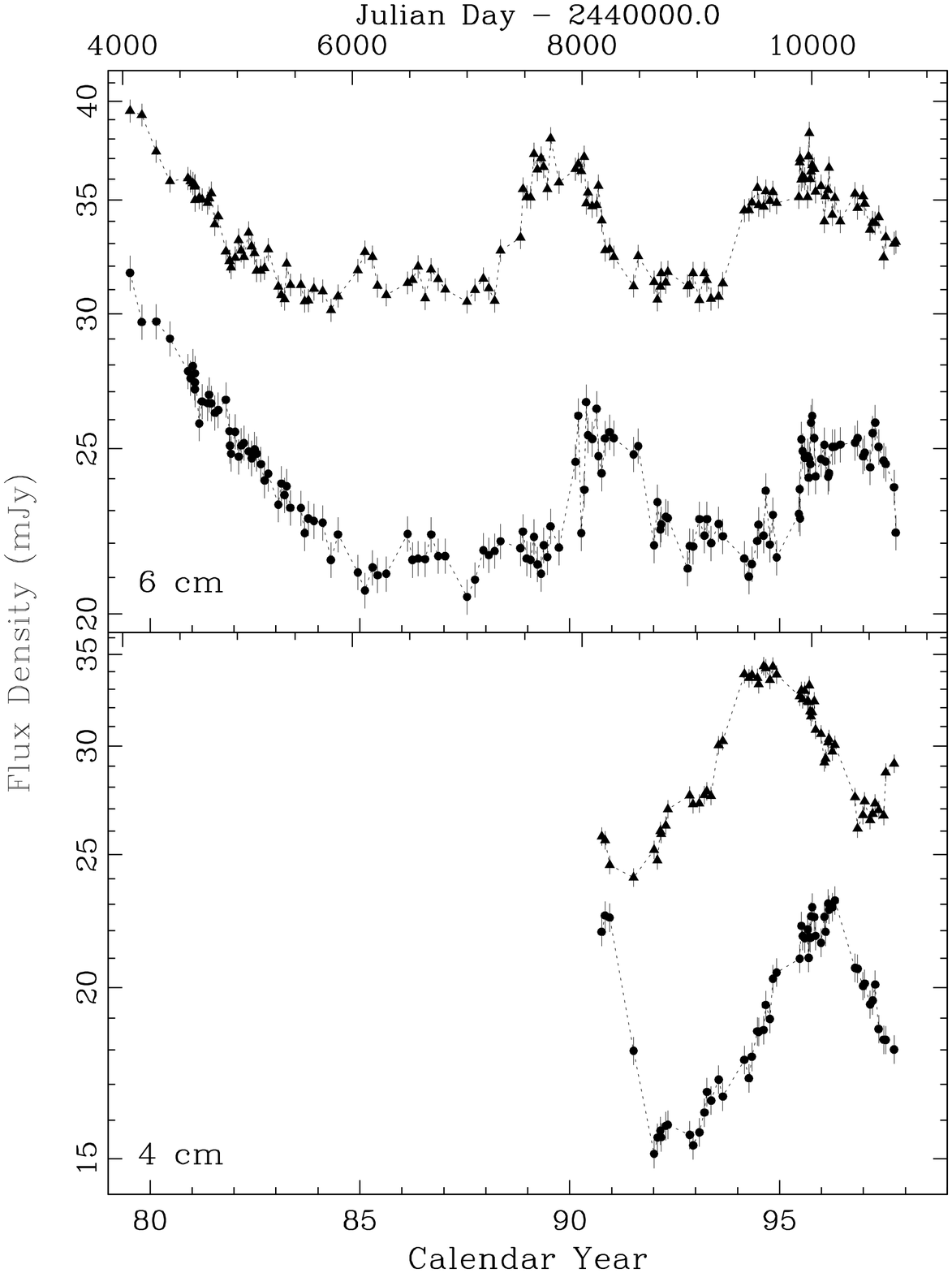}
\caption{The complete 6~cm and 4~cm light curves of gravitational lens
0957+561.  The A image data are shown as triangles and the B image as
circles.  The 4~cm A image has been shifted up by 8\% to avoid overlap
with the B image.}
\end{figure}

\begin{figure}
\plotone{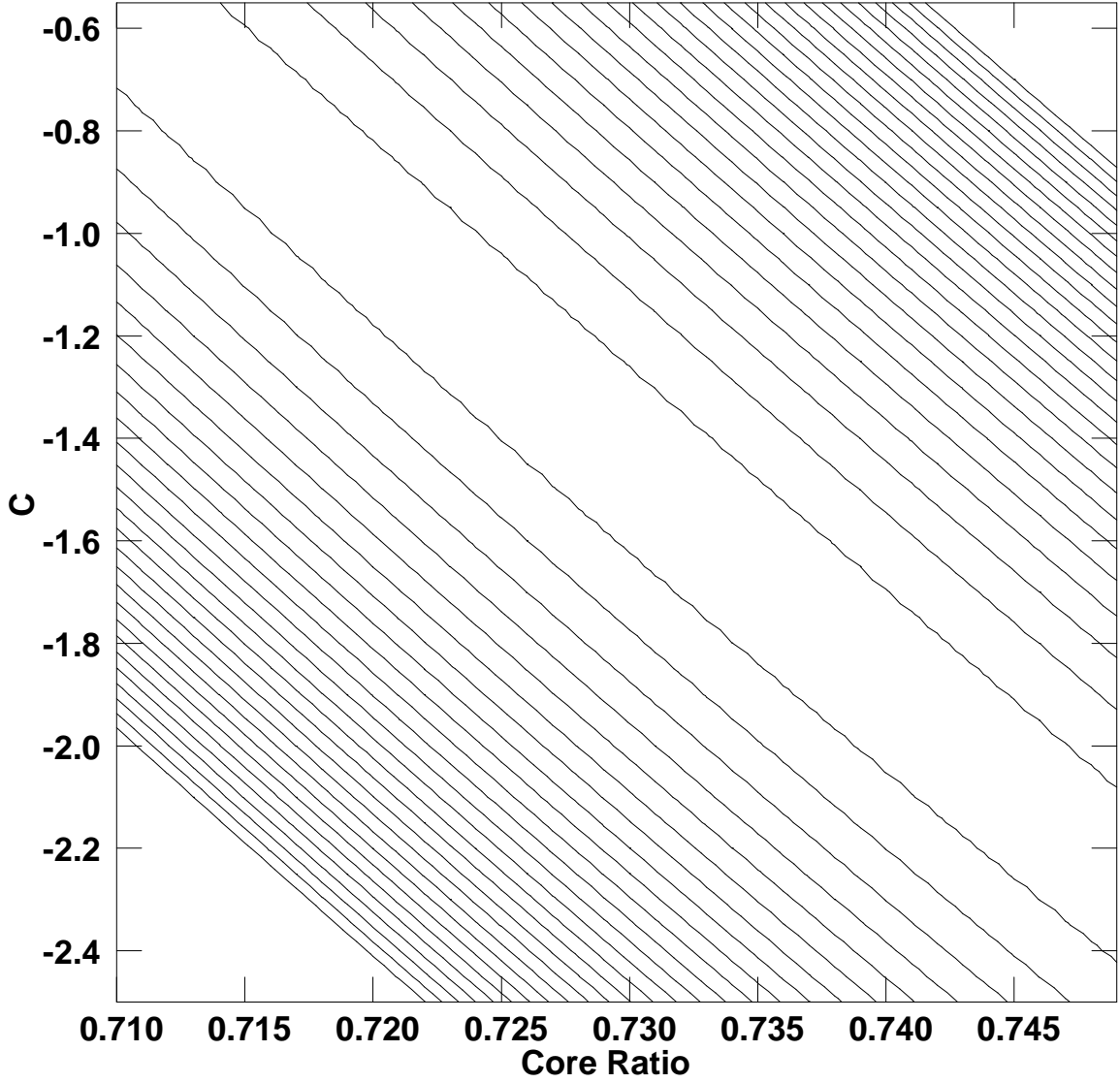}
\caption{PRH$Q$ for the 6$^*$cm light curves, as a function of $c$ and
$\Rcore$.  The delay is fixed at 452~days.  The minimum is $Q=125.6$
at $c=-1.47$ and $\Rcore=0.731$.  Contours start at $Q=130$ and
increase by 10 to $Q=380$.}
\end{figure}

\begin{figure}
\plotone{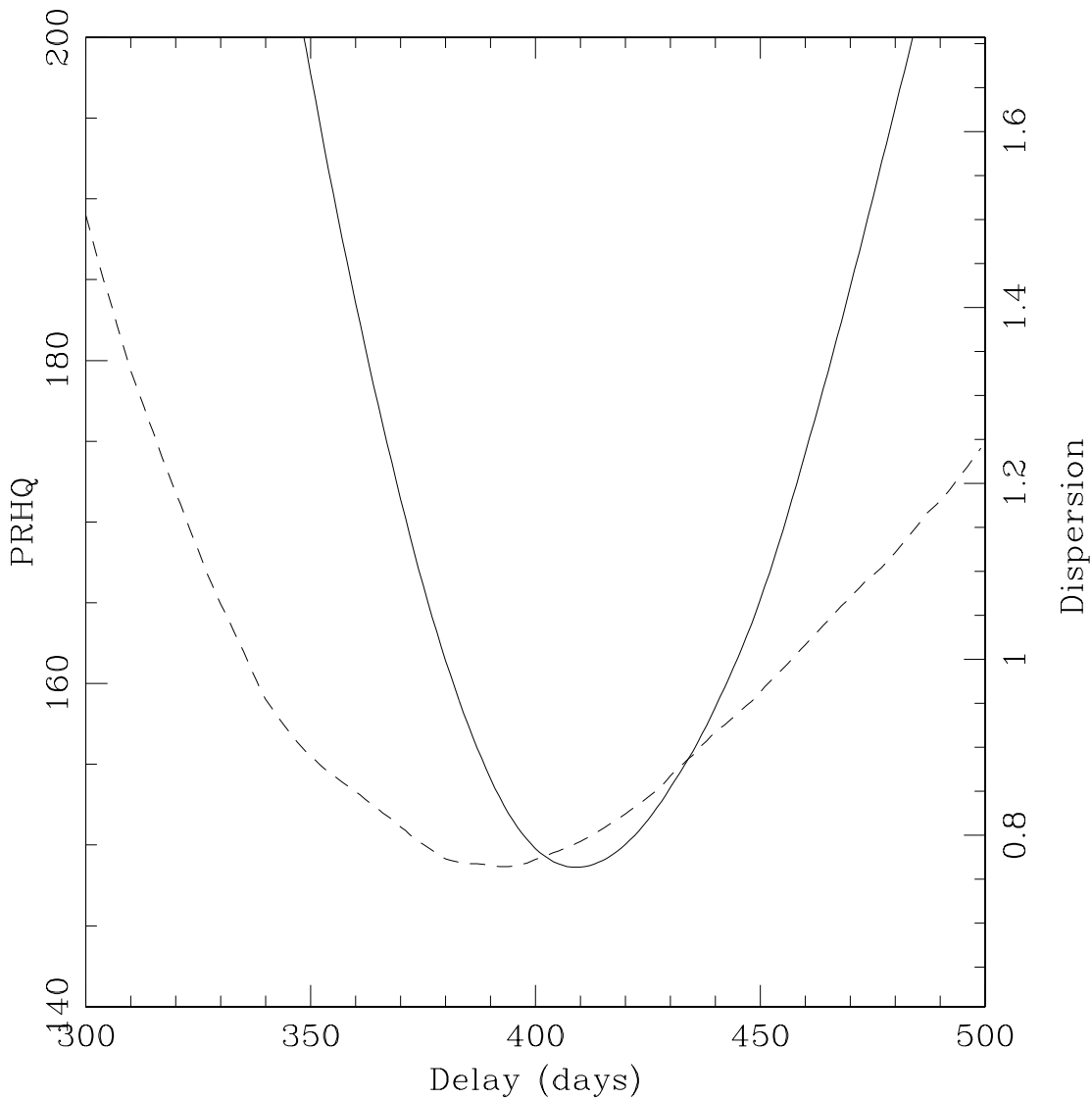}
\caption{Joint time delay analysis of the $6^*$~cm and 4~cm light
curves.  The PRH$Q$ statistic is shown as a solid line with a minimum
at 409~days. The Dispersion statistic is shown as a dashed line with a
minimum at 395~days.  The values of $\Rcore$, $c_6$, and $c_4$ were
set at the best fit values for purposes of plotting the statistic {\it
vs.} delay. The vertical axes were scaled such that the PRH$Q$ and
Dispersion confidence intervals were the same height. }
\end{figure}

\begin{figure}
\plotone{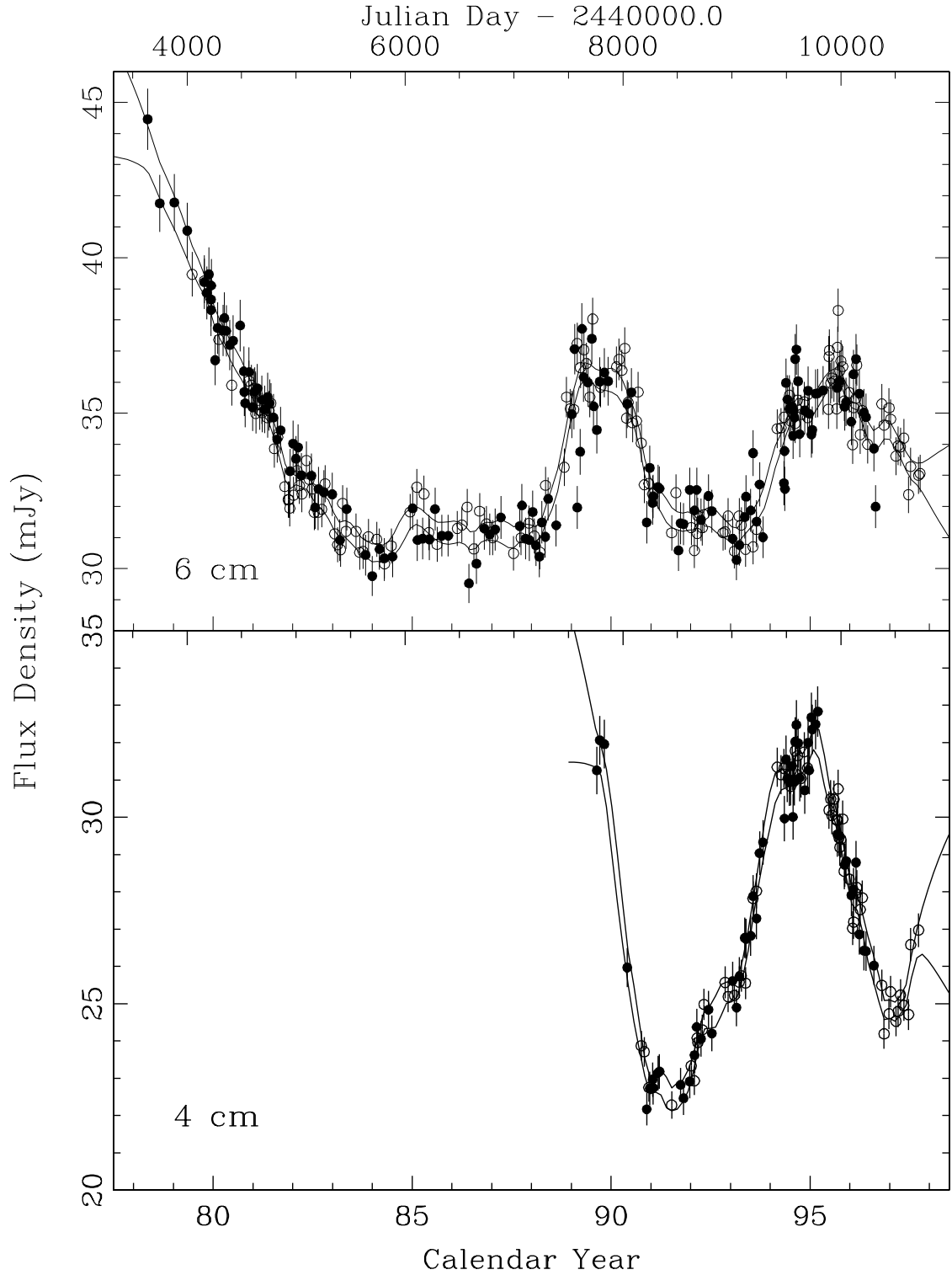}
\caption{The 6~cm and 4~cm light curves combined at at $\tau=409$~days,
$\Rcore=0.753$, $c_6=-2.35$, and $c_4=-2.09$, shifted to the time and
flux density of the A image.  The A image data are shown as open circles
and the B image as solid circles.  The one sigma width of the PRH optimal
reconstruction is shown as a pair of lines.}
\end{figure}
 
%%%%%%%%%%%%%%%%%%%%%%%%%%%%%%%%%%%%%%%%%%%%%%%%%%%%%%%%%%%%%%%%%%%%%%%%%

\begin{deluxetable}{lrccccc}
\tablecolumns{5}
\tablenum{1}
\tablewidth{0pt}
\tablecaption{6~cm Light Curve Data}
\tablehead{
 Calendar & Day\tablenotemark{a}\phm{X} & Array & 
                        \multicolumn{2}{c}{Flux Density (mJy)} \\ \cline{4-5}
 Date     &     &      & A Image    & B Image 
}
\startdata
1995 Jun 18  & 9886.57   &D$\rightarrow$A & 35.13  & 22.89  \nl
1995 Jun 23  & 9892.50   & A              & 36.82  & 23.67  \nl
1995 Jun 28  & 9896.53   & A              & 37.02  & 22.75  \nl
1995 Jul 08  & 9907.23   & A              & 35.97  & 25.32  \nl
1995 Jul 21  & 9919.51   & A              & 36.13  & 24.92  \nl
1995 Aug 07  & 9937.34   & A              & 36.02  & 24.69  \nl
1995 Sep 01  & 9962.33   & A              & 35.14  & 24.75  \nl
1995 Sep 09  & 9970.14   &A$\rightarrow$AnB & 37.12  & 24.04  \nl
1995 Sep 15  & 9976.10   & AnB            & 38.31  & 24.65  \nl
1995 Sep 23  & 9984.17   & AnB            & 35.99  & 24.48  \nl
1995 Sep 30  & 9991.17   & AnB            & 36.36  & 25.90  \nl
1995 Oct 10  & 10001.17  & B              & 36.68  & 26.13  \nl
1995 Oct 27  & 10018.20  & B              & 36.50  & 25.36  \nl
1995 Nov 09  & 10031.06  & B              & 35.39  & 24.08  \nl
1995 Dec 26  & 10077.95  & B              & 35.66  & 24.65  \nl
1996 Jan 26  & 10108.83  & BnC            & 33.99  & 25.13  \nl
1996 Feb 05  & 10118.77  & BnC            & 35.18  & 24.57  \nl
1996 Feb 26  & 10139.67  & C              & 35.48  & 24.07  \nl
1996 Mar 04  & 10146.68  & C              & 36.55  & 24.18  \nl
1996 Apr 05  & 10178.56  & C              & 34.31  & 25.06  \nl
1996 Apr 25  & 10198.66  & C              & 35.09  & 25.07  \nl
1996 Jun 11  & 10246.48  & CnD            & 34.00  & 25.14  \nl
1996 Oct 19  & 10376.05  & A              & 35.30  & 25.20  \nl
1996 Nov 10  & 10397.95  & A              & 34.62  & 25.36  \nl
1996 Dec 26  & 10443.98  & A              & 35.17  & 24.74  \nl
1997 Jan 10  & 10458.89  & A              & 34.81  & 24.86  \nl
1997 Feb 26  & 10505.82  & B              & 33.62  & 24.38  \nl
1997 Mar 19  & 10526.69  & B              & 33.95  & 25.53	\nl
1997 Apr 10  & 10548.65  & B              & 33.92  & 25.90	\nl
1997 May 11  & 10579.65  & B              & 34.20  & 25.06	\nl
1997 Jun 22  & 10622.38  & BnC            & 32.38  & 24.60	\nl
1997 Jul 11  & 10641.43  & C              & 33.28  & 24.49	\nl
1997 Sep 22  & 10714.17  & C              & 33.00  & 23.73  \nl
1997 Oct 06  & 10728.18  & CnD            & 33.06  & 22.32  \nl
\enddata
\tablenotetext{a}{Julian Day -- 2,440,000.0}
\end{deluxetable}

%%%%%%%%%%%%%%%%%%%%%%%%%%%%%%%%%%%%%%%%%%%%%%%%%%%%%%%%%%%%%%%%%%%%%%%%%

\begin{deluxetable}{lrccccc}
\tablecolumns{5}
\tablenum{2}
\tablewidth{0pt}
\tablecaption{4~cm Light Curve Data}
\tablehead{
 Calendar & Day\tablenotemark{a}\phm{X} & Array & 
                        \multicolumn{2}{c}{Flux Density (mJy)} \\ \cline{4-5}
 Date     &     &      & A Image    & B Image 
}
\startdata
1990 Oct 04  & 8169.22   & BnC  & 23.87 & 21.96 \nl 
1990 Nov 01  & 8197.06   & C    & 23.71 & 22.57 \nl
1990 Dec 13  & 8238.89   & C    & 22.75 & 22.49 \nl
1991 Jul 10  & 8448.40   & A    & 22.28 & 17.98 \nl 
1992 Jan 06  & 8627.97   & B    & 23.33 & 15.12 \nl
1992 Feb 04  & 8656.80   & BnC  & 22.93 & 15.54  \nl
1992 Feb 29  & 8681.74   & C    & 24.08 & 15.72 \nl
1992 Mar 07  & 8688.67   & C    & 23.96 & 15.55 \nl
1992 Apr 18  & 8730.60   & C    & 24.30 & 15.85 \nl
1992 May 03  & 8745.60   & C    & 24.98 & 15.88 \nl
1992 Nov 11  & 8938.09   & A    & 25.57 & 15.61 \nl
1992 Dec 10  & 8966.97   & A    & 25.19 & 15.34 \nl
1993 Feb 05  & 9023.78   & AnB  & 25.22 & 15.68 \nl
1993 Mar 21  & 9067.64   & B    & 25.58 & 16.21 \nl
1993 Apr 09  & 9086.67   & B    & 25.75 & 16.78 \nl
1993 May 18  & 9126.48   & B$\rightarrow$BnC & 25.55 & 16.54 \nl
1993 Jul 25  & 9194.21   & C    & 27.82 & 17.13 \nl
1993 Aug 26  & 9226.26   & C    & 28.02 & 16.65 \nl
1994 Mar 04  & 9415.73   & A    & 31.34 & 17.71 \nl
1994 Apr 11  & 9453.68   & A    & 31.14 & 17.17 \nl
1994 May 07  & 9479.63   & A$\rightarrow$AnB & 31.31 & 17.80 \nl
1994 Jun 25  & 9528.52   & B    & 31.15 & 18.58 \nl
1994 Jul 06  & 9540.42   & B    & 30.83 & 18.56 \nl
1994 Aug 18  & 9583.28   & B    & 31.78 & 18.62 \nl
1994 Sep 08  & 9604.27   & B    & 31.65 & 19.42 \nl
1994 Oct 10  & 9636.18   & BnC  & 31.04 & 18.97 \nl
1994 Nov 07  & 9664.08   & C    & 31.75 & 20.29 \nl
1994 Dec 08  & 9694.92   & C    & 31.32 & 20.51 \nl
1995 Jun 23  & 9892.50   & A    & 30.19 & 20.99 \nl
1995 Jul 08  & 9907.23   & A    & 30.49 & 22.18 \nl
1995 Jul 21  & 9919.51   & A    & 30.05 & 21.80 \nl
1995 Aug 07  & 9937.34   & A    & 30.48 & 21.72 \nl
1995 Sep 01  & 9962.33   & A    & 29.94 & 22.05 \nl
1995 Sep 09  & 9970.14   & A$\rightarrow$AnB & 29.91 & 21.02 \nl
1995 Sep 15  & 9976.10   & AnB  & 30.76 & 21.78 \nl
1995 Sep 23  & 9984.17   & AnB  & 29.45 & 21.73 \nl
1995 Sep 30  & 9991.17   & AnB  & 29.19 & 22.54 \nl
1995 Oct 10  & 10001.17  & B    & 29.41 & 22.88 \nl
1995 Oct 27  & 10018.20  & B    & 29.95 & 22.51 \nl
1995 Nov 09  & 10031.06  & B    & 28.55 & 21.81 \nl
1995 Dec 26  & 10077.95  & B    & 28.34 & 21.56 \nl
1996 Jan 26  & 10108.83  & BnC  & 27.02 & 22.52 \nl
1996 Feb 05  & 10118.77  & BnC  & 27.19 & 21.96 \nl
1996 Feb 26  & 10139.67  & C    & 27.95 & 23.03 \nl
1996 Mar 04  & 10146.68  & C    & 28.11 & 22.79 \nl
1996 Apr 05  & 10178.56  & C    & 27.52 & 22.89 \nl
1996 Apr 25  & 10198.66  & C    & 27.84 & 23.15 \nl
1996 Oct 19  & 10376.05  & A    & 25.49 & 20.67 \nl
1996 Nov 10  & 10397.95  & A    & 24.19 & 20.64 \nl
1996 Dec 26  & 10443.98  & A    & 24.73 & 20.05 \nl
1997 Jan 10  & 10458.89  & A    & 25.32 & 20.13 \nl
1997 Feb 26  & 10505.82  & B    & 24.53 & 19.44 \nl
1997 Mar 19  & 10526.69  & B    & 24.79 & 19.56 \nl
1997 Apr 10  & 10548.65  & B    & 25.23 & 20.10 \nl
1997 May 11  & 10579.65  & B    & 24.95 & 18.65 \nl
1997 Jun 22  & 10622.38  & BnC  & 24.71 & 18.32 \nl
1997 Jul 11  & 10641.43  & C    & 26.58 & 18.31 \nl
1997 Sep 22  & 10714.17  & C    & 26.97 & 18.02 \nl
\enddata
\tablenotetext{a}{Julian Day -- 2,440,000.0}
\end{deluxetable}

%%%%%%%%%%%%%%%%%%%%%%%%%%%%%%%%%%%%%%%%%%%%%%%%%%%%%%%%%%%%%%%%%%%%%%%%%

\begin{deluxetable}{llccccccc}
\footnotesize
\tablecolumns{9}
\tablenum{3}
\tablewidth{0pt}
\tablecaption{Results from PRH$Q$ Statistic}
\tablehead{
Light Curve   	& Degrees of  	& $Q$ & \multicolumn{2}{c}{PRH$\chi^2$} & Time Delay		& Core 				& \multicolumn{2}{c}{$c$ (mJy)}	\nl 
		& Freedom     	&	&     	&     	        	& (days)                & Flux Ratio			&               &               \nl \cline{4-5} \cline{8-9}
		&             	&	& 6~cm	& 4~cm	        	&                       & 				& 6~cm          & 4~cm          
}	
\startdata
6 cm          	& 291		& 165.5	& 344	& \nodata		& $459^{+12}_{-15}$	& $0.720^{+0.015}_{-0.014}$	& $-1.04^{+0.58}_{-0.61}$ & \nodata		    \nl \tablevspace{12pt}

6\tablenotemark{*} cm & 283	& 125.6	& 297 	& \nodata		& $452^{+14}_{-15}$	& $0.731^{+0.014}_{-0.014}$	& $-1.47^{+0.56}_{-0.62}$ & \nodata		    \nl \tablevspace{12pt}

4 cm          	& 113		& 15.7	& \nodata & 111 		& $397^{+12}_{-12}$     & $0.762^{+0.026}_{-0.031}$	& \nodata 		  & $-2.44^{+0.94}_{-0.77}$ \nl \tablevspace{12pt}

6 cm \&~4 cm   	& \nodata	& 194.1	& 352 	& 115 			& $416^{+9}_{-8}$     	& $0.744^{+0.011}_{-0.011}$	& $-2.02^{+0.45}_{-0.49}$ & $-1.78^{+0.37}_{-0.32}$ \nl \tablevspace{12pt}

6\tablenotemark{*} cm \&~4 cm & \nodata	& 148.6	& 302 	& 113 		& $409^{+9}_{-9}$ 	& $0.753^{+0.011}_{-0.012}$ 	& $-2.35^{+0.45}_{-0.49}$ & $-2.09^{+0.36}_{-0.33}$ \nl \tablevspace{12pt}
\enddata	
\tablenotetext{*}{6~cm light curve with four points removed, see \S4}
\end{deluxetable}

%%%%%%%%%%%%%%%%%%%%%%%%%%%%%%%%%%%%%%%%%%%%%%%%%%%%%%%%%%%%%%%%%%%%%%%%%
\begin{deluxetable}{lcc}
\tablecolumns{3}
\tablenum{4}
\tablewidth{0pt}
\tablecaption{Results from Dispersion Statistic}
\tablehead{
Light Curve   	& Dispersion	& Time Delay (days)
}	
\startdata
6 cm          	& 0.608		& $430^{+23}_{-22}$	\nl \tablevspace{6pt}

6\tablenotemark{*} cm & 0.489	& $416^{+22}_{-24}$	\nl \tablevspace{6pt}

4 cm          	& 0.259		& $383^{+15}_{-19}$    \nl \tablevspace{6pt}

6 cm \&~4 cm   	& 0.915		& $397^{+12}_{-15}$     \nl \tablevspace{6pt}

6\tablenotemark{*} cm \& 4 cm & 0.764	 & $395^{+13}_{-15}$     \nl \tablevspace{6pt}
\enddata	
\tablenotetext{*}{6~cm light curve with four points removed, see \S4}
\end{deluxetable}

%%%%%%%%%%%%%%%%%%%%%%%%%%%%%%%%%%%%%%%%%%%%%%%%%%%%%%%%%%%%%%%%%%%%%%%%%

\end{document}